\documentstyle[eqsecnum,multicol,aps,psfig]{revtex}

\draft

\headheight=\baselineskip
\begin{document}
\input{epsf}

\title{
       \vspace*{-0.5cm}\hfill {\tt cond-mat/9506016}
       \vspace*{0.5cm}
       \\
Fracturing of brittle homogeneous solids. \\
Finite-size scalings.
}

\author{Frank Tzschichholz}

\address{P.M.M.H. (CNRS, URA 857), \'Ecole Sup\'erieure de Physique
et de Chimie Industrielle de la ville de Paris,\\
10 rue Vauquelin, 75231 Paris Cedex 05, France~~\cite{email}}

\date{\today}
\maketitle

\begin{abstract}
Using a two dimensional lattice model we investigate the crack
growth under the influence of remote tensile forces as well as due
to an internally applied pressure (hydraulic fracturing).
For homogeneous elastic properties
we present numerical finite-size scalings for the breaking stresses
and pressures in terms of crack lengths and lattice sizes.
Continuum theory predicts for the tensile and for the pressure problem
identical scaling functions. Our findings for
the tensile problem are in very good agreement with continuum results.
However, for the hydraulic fracture problem we observe a
different finite-size scaling.
We explicitly demonstrate that the modified scaling is a consequence
of the discrete structure of the lattice (micro structure).

\end{abstract}


\newsavebox{\boxl}
\savebox{\boxl}{$\ell$}

\begin{multicols}{2}
\section{Introduction}

About 30 years ago considerable interest has been devoted to
the properties of generalized Cosserat elasticity.
One of the main
activities was the formulation of continuum theories
reproducing lattice dispersion relationships to a certain
extent\cite{b3-4b}. These approaches (gradient theories) were
designed
in order to incorporate certain atomistic aspects into  generalized
elastic continua. However, it was early recognized that such
theories were in principle also applicable to the description of
micro structural and dislocation phenomena\cite{b3-4c}.
Recently a first order (strain) gradient theory was
applied to a fracture mechanical problem
obtaining results reminiscent of
Barrenblatt's cohesion theory\cite{b3-4d,b3-4e}.
However, the gradient elasticity approach
leads to, in general, formidable mathematical difficulties, to a
somewhat
suspect choice of additional boundary conditions, and to numerous
physically not well understood new material constants.

These difficulties are circumvented in discrete fracture models
which have been adopted during the past decade in order to study
the effect of structural heterogeneities and fluctuations on
fracture growth processes under various physical conditions.
Using triangular networks of springs stretched on the
outer boundary into the six directions of a hexagon the
breaking of a material from a central hole has been investigated
by several authors \cite{b3}. Employing disordered beam networks,
the scaling of micro crack densities \cite{b3-2} and
the growth of cracks due to an inner hydraulic pressure \cite{b3-3}
have been studied in some detail.
These investigations offer
a new methodological approach to study failure processes
in heterogeneous brittle materials
on a mesoscopic length scale
close to experimental situations \cite{b3-4}.
Our interest is focused on brittle materials because their
rheological
behavior can be regarded most simple
as they do not show any
remanent (plastic) deformations\cite{b3-4a}.

One difficulty arising during the examination of numerically
observed
results often consists in finding appropriate scaling variables.
As we will see below this is not a trivial
task, not even for perfect straight cracks. We focus
on finite-size scaling
because the employed lattice sizes $L$ are always finite
one is, however, mostly interested in properties of very large or
infinite systems. Thus the finite-size scaling approach
appears to be the only conclusive numerical method to verify
(conjectured) asymptotic behavior for $L\to\infty$.

In the present paper
we will show that the micro structure (lattice structure) can
change both asymptotic and finite-size scaling of the
breaking characteristics, depending on the employed boundary
conditions.

In Sec.~\ref{sec:Model} we briefly outline the fracture model,
the employed
boundary conditions and how to determine the
external breaking stresses and pressures.
Following this we present in Sec.~\ref{sec:Results} the main
finite-size scaling results
for breaking stresses and pressures of linear cracks
on two-dimensional beam lattices. We compare our findings
with known analytical continuum results.

For crack growth
scaling properties in disordered materials
the reader is referred to Ref.~\ref{b3-2}.

\section{The Model}
\label{sec:Model}

We investigate both the case of uniaxial
loading due to remote tensile forces and
the loading due to a
hydrostatic pressure acting from the interior on the crack surface.
The former situation is typically encountered when a material is
pulled at its outer faces, while the latter one can be
found in engineering applications of hydraulic
fracturing\cite{b3-2a}.
We consider the beam model
on a two dimensional square lattice of linear
size $L$ \cite{b3-4,b3-5}. Each of the
lattice sites $i$ carries three real variables:
the two translational displacements $x_i$ and $y_i$ and
a rotational angle $\varphi_i $. Neighboring sites
are rigidly connected by elastic beams of length $l$.
The beams all have the same cross section  and
the same elastic behavior, governed by three material
dependent constants $a=l/(EA)$, $b=l/(GA)$
and $c=l^3/(EI)$ where
$E$ and $G$ are the Young and shear moduli, $A$ the cross
section of
the beam and $I$ the moment of inertia for flexion.
We used for all simulations the values $a=1.0$, $b=0.0017$
and $c=8.6$.
When a site is rotated ($\varphi_i \ne 0$)
the beams bend accordingly always forming
tangentially right angles with each other.
In this way local momenta are taken into account.
For a horizontal beam between sites $i$
and $j$ one has the longitudinal force acting at site $j$:
$F_{ij} = \alpha (x_i -x_j)$;
the shear force:
$S_{ij} = \beta (y_i - y_j) +{\beta\over 2}l
(\varphi_i +\varphi_j)$,
and the flexural torque at site $j$:
$M_{ij} ={\beta\over 2}l(y_i -y_j +l\varphi_j) +
\delta l^2(\varphi_i-\varphi_j)$, using $\alpha = 1/a$,
$\beta=1/(b+c/12)$ and $\delta=\beta(b/c+1/3)$.
Corresponding equations hold for the vertical beams.
The hydrostatic stress acting on a beam is given by its
longitudinal force per beam cross section, $\sigma_{ij}=F_{ij}/A$.
In  mechanical equilibrium the sum over all internal and
external forces (torques) acting on site $j$ must vanish.
We will neglect inertial forces, assuming a
sufficiently slow fracture process.
The cohesive properties of the beams are considered to be
homogeneous, i.e.\
the beams behave linear-elastic up to their theoretical strength
$F_{th}$ (cohesion force), above which they irreversibly
break (zero elastic constants).
At the beginning of each simulation we break one vertical beam
located at the center of the lattice. This is the initial crack.
Corresponding to the employed boundary conditions described below
we calculate the internal force distribution $F_{ij}$ of beams
connecting sites $i$ and $j$ using a conjugate gradient
method\cite{b3-6}. From this longitudinal forces we
determine and break the most over stressed beam, i.e.
the beam carrying the highest force,
$F_{i_0j_0}= \max_{\{ij\}}F_{ij}$. The maximum force
allows us to calculate a scaling factor $\lambda = F_{th} /
F_{i_0j_0}=\sigma_{th}/\sigma_{i_0j_0}$ which in
turn is used to determine the
macroscopic breaking stress $\sigma_c$
or breaking pressure $P_c$ (measured in units of $\sigma_{th}$)
necessary to fulfill the
local fracture criterion
$\sigma_{i_0j_0}=\sigma_{th}$.
Such scaling is possible due to the linearity of the employed
elastic equations.
Breaking the beam connecting sites $i_0$ and $j_0$ destroys
the balance of forces at those sites and one has to relax
the system to its new equilibrium configuration. Then the
above steps are repeated until the lattice breaks apart.

It is well known that the highest
stress enhancement factors occur at the crack tips
for the pressure as well as
for the tensile problem. Because we consider only
homogeneous cohesion forces the terms `most over stressed' and
`highest stress' are equivalent and the cracks grow linear
in direction of the highest local tensile force. In our case
the direction of crack growth is
the x-direction because of the uniaxiality
of imposed boundary conditions, see below.

\subsection{Tensile Fracturing}
\label{sec:Tensile-Model}
To impose an external strain we attach at the bottom of the
lattice a
zeroth line on which for all sites $x_i=y_i=\varphi_i=0$
are fixed,
and on the top we attach a $(L+1)$st line on which all sites
have the same fixed values $x_i=\varphi_i=0$ and $y_i=1$.
With this boundary conditions the external displacement in
y-direction
is fixed to unity (Dirichlet boundary conditions) and one can
imagine them as being represented by rigid bars attached
at the top and the bottom of the lattice\cite{b3-5}.
For the boundary conditions in x-direction we consider
periodic as well as free lattice boundaries.
The externally applied stress, $\sigma_{0}$, necessary to maintain
a unity
displacement between the two rigid bars can be easily calculated
by summing up all $y_i$ displacements over the first line,
$\sigma_0 = {\alpha\over{A\,L}}\sum y_i$, where $\alpha$ is the
longitudinal force constant, $A$ the cross section of the beam and
$L$ the (dimensionless) lattice size. Employing the above defined
scaling factor $\lambda$ we express the externally applied
breaking stress, $\sigma_c$, in the form,
\begin{equation}\label{eq:Tensile}
\sigma_c = \lambda \sigma_0,
\end{equation}
or in dimensionless form,
\begin{equation}\label{eq:Tensile-Stress}
{\sigma_c\over\sigma_{th}}= {1\over F_{i_0j_0}}
{\alpha\over L}\sum y_i.
\end{equation}
With this notation the external breaking stress $\sigma_c$
is just the
stress necessary to yield the local breaking condition
$\sigma_{i_0j_0}=\sigma_{th}$.

We will use Eq.(\ref{eq:Tensile-Stress}) for our
finite-size scaling purposes.
It expresses the breaking stress,
measured in units of cohesion
stress, in terms of the beam force $F_{i_0j_0}$ at the `crack
tip' and
the average force ${\alpha\over L}\sum y_i $ on a beam due to
the global unit displacement.
In general both terms depend on the number of broken beams $N$
(crack size) and on the lattice size $L$,
see Sec.~\ref{sec:Results-Tensile}.

\subsection{Hydraulic Fracturing}
\label{sec:Hydraulic-Model}
Experimentally hydraulic fracturing is encountered
when an incompressible fluid
is injected under high pressure into an existing crack.
The characteristics of this particular loading mode is that
the loading of the crack only happens on the crack surface
itself, but not
on remote surfaces like in the case of tensile fracturing.
Recently hydraulic fracturing has been numerically addressed
by introducing for each
broken beam a force dipole of strength $F_0$ simulating
an inner pressure $P_0= F_0/ A$\cite{b3-3}.
In the present paper we will follow this implementation.
Starting from
one vertical broken beam (zero elastic constants, one
vertical force dipole)
we calculate the internal stress distribution $F_{ij}$ using
the conjugate
gradient. Calculating the scaling factor $\lambda$ as
described above
we determine and break the beam carrying the highest stress,
$F_{i_0j_0}$.
Thereafter the lattice has to be relaxed again
(now two broken beams, two force dipoles) and one repeats the above
mentioned steps until the lattice breaks apart.

We have performed calculations for free external boundaries
in x- and y-direction
as well as for periodic boundaries in
x-direction and free boundaries in y-direction.
As we only impose stresses the elastic
solution (displacement field) is unique up to a general translation
and rotation. Without restriction we therefore fix
the displacement of an arbitrary lattice site to zero.
Analogous to Eqs.~(\ref{eq:Tensile})and (\ref{eq:Tensile-Stress})
one has for the breaking pressure,

\begin{equation}\label{eq:Pressure}
P_c = \lambda P_0,
\end{equation}
and
\begin{equation}\label{eq:Pressure-Stress}
{P_c\over \sigma_{th}}={F_0\over F_{i_0j_0}},
\end{equation}
respectively.
It should be noted that $\sigma_{th}$ is
the tensile strength
and not the compressional strength.
In fact the stresses at the tips are tensile stresses as the crack
is hydraulically opened.

We will use Eq.~(\ref{eq:Pressure-Stress})
for the finite-size scaling of the hydraulic problem.

\section{Results}
\label{sec:Results}
As explained in Sec.~\ref{sec:Model} we obtain from our simulations
the macroscopic breaking stresses and pressures,
$\sigma_c(N,L)$ and $P_c(N,L)$, measured in units of the theoretical
strength $\sigma_{th}$,
in terms of the number of broken
beams $N$ and the employed lattice sizes $L$.
As we are interested in the finite-size scaling properties of the
breaking stresses ($\Sigma_c =\sigma_c$) and pressures ($\Sigma_c =P_c$)
we introduce two new functions, $\Psi (N)$ and
$\Phi (N/L)$,
\begin{equation}\label{eq:Scaling-General}
\Sigma_c(N,L)=\sigma_{th}\,\Psi (N)\,\Phi (N/L).
\end{equation}

The function $\Phi (N/L)$ describes the finite-size
corrections to the scaling $\Psi (N)$ between the breaking
stress/pressure
and the crack length for given boundary conditions in the infinite
system. The quantity $N/L$ is for our purposes the most suitable
scaling variable as it represents the length of a linear crack
measured in units of the employed lattice size.
In general both functions, $\Psi$ and $\Phi$ will depend on the
employed boundary conditions.
In order to show numerically that the finite-size scaling
function $\Phi (N/L)$ exists, one needs to know (or to
conjecture) the
explicit form of $\Psi (N)$, i.e. the asymptotic
behavior of $\Sigma_c$ as $L\to\infty$.

We will access the latter information from simple continuum
mechanical considerations of symmetric elasticity.

\subsection{Tensile Fracturing}
\label{sec:Results-Tensile}
\begin{figure}[htb]
\centerline{
        \epsfxsize=8.0cm
        \epsfbox{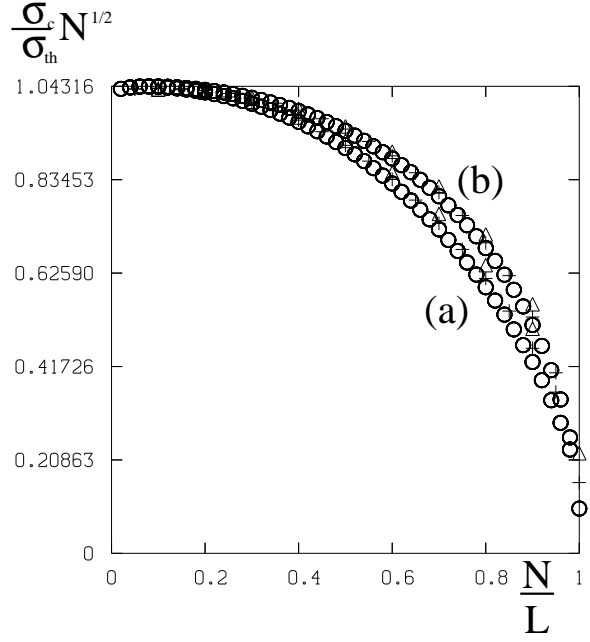}
        \vspace*{0.5cm}
        }
\caption{
         Finite-size scaling for the rescaled breaking stress
         $\sigma_c/\sigma_{th}N^{1/2}$ as a function of the
         rescaled crack length $N/L$, (a) for free boundaries in
         x-direction and (b) for periodic boundaries in x-direction.
         Lattice sizes: ($\triangle$) for $L=20$, ($+$) for $L=40$
         and ($\circ$) for $L=100$.
        }
\label{fig:Tension-Scaling}
\end{figure}

Previous continuum mechanical scaling considerations have
successfully been applied to fracture models.
It has been shown \cite{b3-7} that the breaking stress $\sigma_c$
for tensile fracturing of a finite, periodic beam lattice follows
very well
the known `Tangent-Formula' of fracture mechanics
\cite{b3-8,b3-9},
\begin{equation}\label{eq:Tangent-Formula}
{\sigma_c\over\sigma_{th}}\sim
N^{-1/2}\sqrt{ {N\over L}\cot{{\pi\over 2}{N\over L}}}.
\end{equation}
By comparison with Eq.~(\ref{eq:Scaling-General}) we see that
the asymptotic scaling for the breaking stress $\sigma_c$ is given
by $\Psi\sim N^{-1/2}$, a result that originally was
found by Griffith.
In Fig.~\ref{fig:Tension-Scaling} we show the finite-size scalings
for the numerically obtained tensile breaking stresses
(a) for free boundary conditions and
(b) for periodic boundary conditions.
For both curves the data collapse is quite acceptable. While for
the periodic problem the finite-size correction $\Phi (N/L)$
follows Eq.~(\ref{eq:Tangent-Formula})\cite{b3-7},
no analytical expression for $\Phi$ is known in the case of
free boundary conditions.
However, the differences for the breaking stresses at given values of
$N/L$ are less than $0.1\sigma_{th}$, see
Fig.~\ref{fig:Tension-Scaling}, and Eq.~(\ref{eq:Tangent-Formula})
represents a reasonably good approximation for finite solids for many
practical purposes.

\subsection{Hydraulic Fracturing}
\label{sec:Results-Pressure}
In general analytical solutions of crack problems are
difficult to obtain and even for two-dimensional
problems in infinite, semi-infinite or
periodic domains one is often faced with complicated integral
equations. The reader interested in this is referred to
Muskhelishvili\cite{b3-10} and Sneddon\cite{b3-9}.

\begin{figure}[htb]
\centerline{
        \epsfxsize=8.0cm
        \epsfbox{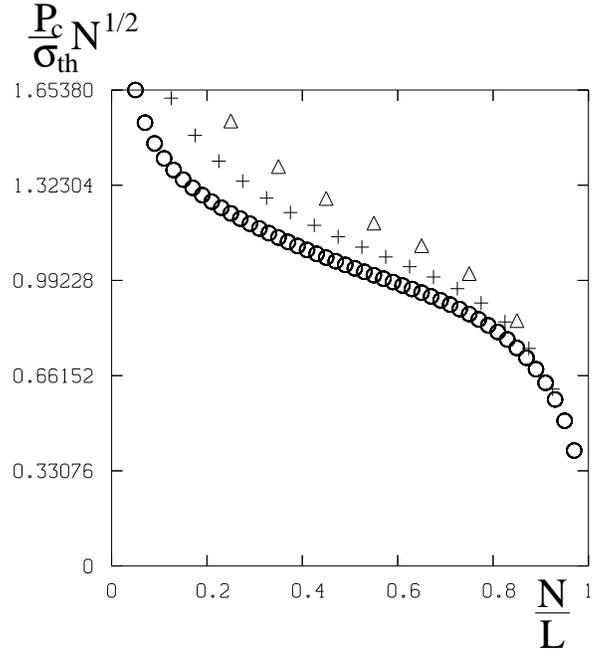}
        \vspace*{0.5cm}
        }
\caption{
         Plot of the rescaled breaking pressure
         $P_c/\sigma_{th}N^{1/2}$ as a function of the
         rescaled crack length $N/L$ for periodic boundaries
         in x-direction. In contrast to
         Fig.~\ref{fig:Tension-Scaling} we do not find any
         data collapse, indicating that $\Psi (N)\sim N^{-1/2}$
         in Eq.~(\ref{eq:Scaling-General}) is not the
         appropriate asymptotic expression for the breaking pressure.
         Lattice sizes: ($\triangle$) for $L=20$, ($+$) for $L=40$
         and ($\circ$) for $L=100$.
        }
\label{fig:Pressure-wrong-Scaling}
\end{figure}

It has been shown analytically that the breaking pressure $P_c$
of a semi-infinite two-dimensional continuum with periodic boundary
conditions is given by Eq.~(\ref{eq:Tangent-Formula})
where one has to replace $\sigma_c$ by $P_c$ \cite{b3-9}.
We show in Fig.~\ref{fig:Pressure-wrong-Scaling} the analogous plot
to Fig.~\ref{fig:Tension-Scaling} for the breaking
pressure. One does not obtain a data collapse, indicating that
$\Psi\sim N^{-1/2}$, see Eqs.~(\ref{eq:Scaling-General}) and
(\ref{eq:Tangent-Formula}), is
{\em not} the scaling of the breaking pressure in the asymptotic limit
of infinite lattices.

This result is somewhat unexpected, because the continuum limit
predicts identical scaling relations for breaking stresses and
pressures and on the other hand we find
good agreement between lattice and continuum scalings
for the tensile problem.

However, the proposed model for hydraulic fracturing is defined on
a lattice and not in a continuum. As lattices always show additional
structure, higher order gradient terms of the
continuum displacement field
appear in the elastic solution, which have no counterparts in
simple continua\cite{b3-11}.

In the following we present an argument why the lattice
finite-size scalings
for the pressure and tensile problem are different.

In the tensile problem the loading forces are acting at the
lattice boundaries {\em remote} to the stress free crack surfaces.
Therefore local stresses close to the crack tips are carried
through nearly the whole elastic volume.
Contrary to this in the hydraulic problem loading
only happens due to force
dipoles acting at the crack surfaces {\em close} to the crack tips.
One therefore might expect that the breaking pressures $P_c$ show a more
pronounced `(micro structure) lattice behavior' than the breaking
stresses $\sigma_c$. For very large cracks these differences should
diminish as the discrete loading approaches a force density.

In order to extent a linear crack of length $N$ in an infinite continuum
one has to impose a breaking pressure $P_c \sim N^{-1/2}$.
However, employing a {\em single double force} of magnitude $F_c$
one finds, $F_c\sim N^{1/2}$ for the critical loading force
(related Boussinesq problem)\cite{b3-12}.

We argue now that for a single crack in an infinite disk loaded by
equidistant force dipoles the asymptotic scaling for $\Psi$
is equivalent to that of an infinite lattice.
Knowing $\Psi$ we plot the finite-size scalings
for the investigated lattices.

The asymptotic continuum scaling for the breaking pressure (as
calculated from equidistant force dipoles)
is most easily obtained using the complex stress
function of Westergaard\cite{b3-12,b3-8}.
We find for the asymptotic scaling in an infinite disk,
\begin{equation}\label{eq:improved-scaling}
\Psi (N)\approx {\pi\over 2}{N^{1/2}\over 1+N\,f(N)},
\end{equation}
with
\begin{equation}
f(N)=\sum_{k=1}^{(N-1)/2}{1\over\sqrt{({N+1\over 2})^2 -k^2}},
\quad\mbox{$N$ odd}.
\end{equation}
The sum $f(N)$ converges rapidly towards $\arcsin({N-1\over N+1})$
and for large $N$ towards $\pi /2$.
Hence in the limit of a large number of force dipoles we obtain
the aforementioned asymptotic Griffith scaling for the breaking pressure,
$P_c / \sigma_{th}=  N^{-1/2}$.

\begin{figure}[htb]
\centerline{
        \epsfxsize=8.0cm
        \epsfbox{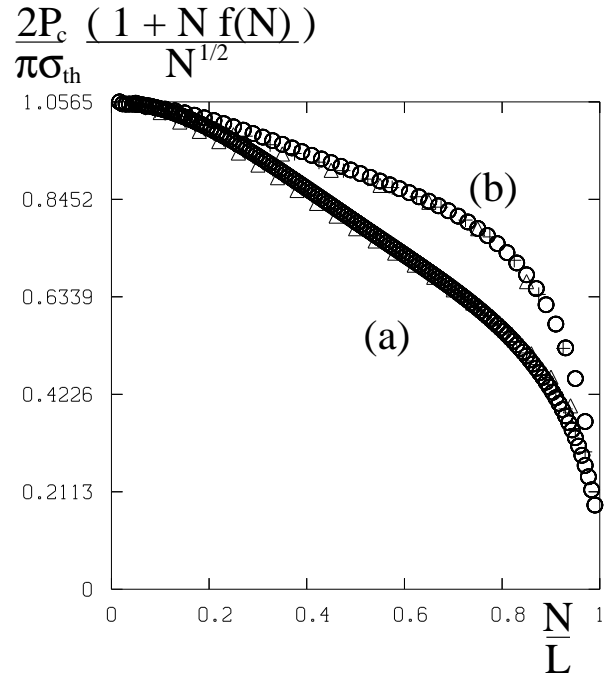}
        \vspace*{0.5cm}
        }
\caption{
         Finite-size scaling for the rescaled breaking pressure
         $P_c/\sigma_{th}\Psi^{-1}(N)$ with
         $\Psi^{-1}(N)={2\over\pi}{1+N\,f(N)\over N^{1/2}}$, see
         Eq.(\ref{eq:improved-scaling}), as a function of the
         rescaled crack length $N/L$;
         (a) for free boundaries $(\triangle):\, L=50$, $(+):\, L=100$,
         $(\circ):\, L=300$; (b) for periodic boundaries
         $(\triangle):\, L=20$, $(+):\, L=40$, $(\circ):\, L=100$.
         Note the data collapses.
        }
\label{fig:Pressure-Scaling}
\end{figure}

In Fig.~\ref{fig:Pressure-Scaling} we show the resulting
finite-size scalings for the breaking pressure, based on
Eqs.~(\ref{eq:Scaling-General}) and (\ref{eq:improved-scaling}).
In this plot a reasonable scaling is obtained, which proves our
earlier statement that the asymptotic limits ($L\to\infty$) of the
breaking pressures for continua and lattices
are different.
Comparing Fig.~\ref{fig:Pressure-Scaling} with
Fig.~\ref{fig:Tension-Scaling} one also notes a significant
deviation from the scaling form Eq.~(\ref{eq:Tangent-Formula}),
demonstrating a modified correction to scaling behavior.

\section{Conclusions}
\label{sec:Conclusions}
We have presented finite-size scaling results for breaking stresses
(tensile fracturing) and breaking pressures (hydraulic fracturing)
of elastically and cohesively homogeneous beam lattices
in two dimensions (plane stress condition).
For the tensile problem it has been shown that
both the asymptotic lattice scaling $\Psi (N)$ as well as
the finite-size lattice scaling $\Phi (N/L)$ are well
represented by the known continuum scaling form
(`Tangent-Formula').

In contrast the hydraulic fracture problem formulated
on a lattice, has lead us to different asymptotic and finite-size
lattice scalings, though the corresponding continuum scalings
for breaking pressures and stresses are identical.
Explicitly we constructed the asymptotic lattice scaling of
breaking pressures from purely continuum mechanical considerations,
by taking into account the discrete nature of the loading.

As the overwhelming number of brittle materials is granular in
character, the interesting question arises whether their
breaking characteristics can be  satisfactory described
by continuum models regardless of employed loading conditions.

\section*{Acknowledgments}
I would like to thank S. Schwarzer for interesting discussions.

%
%

\bibliographystyle{prsty,unsrt}

\end{multicols}
%
%


\end{document}